\documentclass[useAMS]{mn2e}

\def\mnras{MNRAS}
\def\apj{ApJ}
\def\apjl{ApJL}
\def\apjs{ApJS}
\def\aap{A{\&}A}
\def\aapr{A{\&}A Rev.}
\def\aj{AJ}
\def\pasp{PASP}
\def\pasj{PASJ}
\def\nat{Nature}
\def\araa{ARAA}
\usepackage{times}
\usepackage{amssymb}
\usepackage{rotating}
\input{epsf}

\title[The Mass and Spin of 1H 0707-495]
{The Mass and Spin  
of The Extreme Narrow Line Seyfert 1 Galaxy 1H 0707-495 and Its
Implications for The Trigger for Relativistic Jets}

\author[C. Done \& C. Jin]
{Chris Done$\thanks{E-mail:chris.done@durham.ac.uk}^1$, 
  Chichuan Jin$^{2}$
\\
$^1$ Centre for Extragalactic Astronomy, 
Department of Physics, University of Durham, South Road,
Durham DH1 3LE, UK\\
$^2$ Max-Planck-Institut f{\"u}r Extraterrestrische Physik, Giessenbachstrasse, D-85748 Garching, Germany\\
}

\date{Submitted to MNRAS}
\pagerange{\pageref{firstpage}--\pageref{lastpage}} \pubyear{2015}

\begin{document}

\topmargin = -0.5cm

\maketitle

\label{firstpage}

\begin{abstract}
  Relativistic reflection models of the X-ray spectrum of the
  `complex' Narrow Line Seyfert 1 (NLS1) 1H 0707-495 require a high
  spin, moderate inclination, low mass black hole. With these
  parameters fixed, the observed optical/UV emission directly
  determines the mass accretion rate through the outer disc and hence
  predicts the bolometric luminosity.  This is $140-260 \times$ the Eddington
  limit. Such a disc should power a strong wind, and winds are
  generically expected to be clumpy. Changing inclination angle with
  respect to a clumpy wind structure gives a possible explanation for
  the otherwise puzzling difference between `complex' NLS1 such as
  1H 0707-495 and `simple' ones like PG 1244+026. Lines of sight which
  intercept the wind show deep absorption features at iron from
  the hot phase of the wind, together with stochastic dips and complex
  absorption when the clumps occult the X-ray source (complex NLS1),
  whereas both these features are absent for more face-on inclination
  (simple NLS1). This geometry is quite different to the clean view of
  a flat disc which is assumed for the spin measurements in
  relativistic reflection models, so it is possible that even
  1H 0707-495 has low spin. If so, this re-opens the simplest and hence
  very attractive possibility that high black hole spin is a necessary
  and sufficient condition to trigger highly relativistic (bulk
  Lorentz factor $\sim 10-15$) jets.

\end{abstract}


\section{Introduction} \label{sec:introduction}

There are a multiplicity of different types of Active Galactic Nuclei
(AGN), yet there are only a limited number of fundamental parameters
which can determine their behaviours.  There is black hole mass, mass
accretion rate and spin, with inclination, abundances and environment
as further factors which can affect how the AGN is seen as well as
non-deterministic factors such the history of the accretion flow
(hysteresis and/or magnetic flux accumulation).

Black hole mass can now be estimated e.g. through scaling relations
using the FWHM of the H$\beta$ line together with optical
luminosity. A thin disc model fit to the optical/UV data then directly
gives the absolute accretion rate, $\dot{M}$, though the outer disc
for a given inclination angle (Laor \& Davis 2011).  Emission from the
inner disc does not affect this as smaller radii emit at higher
temperatures, so contribute to the far UV rather than to the optical
for typical AGN parameters. Thus simply observing the optical
emission from the outer disc is sufficient to set the absolute mass
accretion rate of the entire flow in standard disc models. 

The inner disc properties instead determine the total
bolometric luminosity as spin sets the inner edge of the disc, and hence the
total efficiency. High prograde spin leads to additional luminosity at higher temperatures
from the same mass accretion rate through the outer disc compared to a lower spin. 
If the
mass, inclination and spin are known, then the absolute $\dot{M}$
through the outer disc directly predicts the total bolometric
luminosity, and hence $L_{bol}/L_{Edd}$ of the accretion
flow. 

Laor \& Davis (2011) used this argument in reverse. They measured
$\dot{M}$ from the optical/UV flux, then integrated the observed
spectral energy distribution (SED) to estimate $L_{bol}$ and hence
constrained spin in their sample of AGN.  However, there are other
ways to constrain spin. Firstly, via the peak temperature and
luminosity of the disc continuum as discussed above. This disc
continuum fitting technique can be used to good effect in black hole binaries,
where the disc peaks at X-ray temperatures (e.g. Ebisawa et al. 1991;
Kubota, Makishima \& Ebisawa 2001; Done, Gierlinski \& Kubota
2007). However, in AGN the disc typically is expected to peak in the
far UV, which is unobservable due to interstellar absorption. Instead,
the iron line profile can be used to constrain spin. X-ray
illumination of the disc produces a fluorescence iron K$\alpha$ line,
whose profile contains the imprint of the (spin dependent) innermost
disc radius as well as the illumination pattern and inclination
(Fabian et al. 1989; 2000). The new reverberation technique uses this
together with the light travel time delays (Fabian et al. 2009) to map
the reflecting structures (see e.g the review by Uttley et al. 2014).

Multiple AGN now have high spin as determined by the iron line profile
(see e.g. the compilation of Reynolds 2013). The most convincing of
these have reverberation mapping as well, and tend to be Narrow Line
Seyfert 1s (NLS1) such as 1H0707-495 (Fabian et al. 2009).  The iron
line profile also gives inclination, so for these systems the
optical/UV continuum gives absolute $\dot{M}$, which, combined with
the known spin gives $L_{bol}$ and hence
$L_{bol}/L_{Edd}$.

Here we use this technique on two specific, well studied NLS1, PG
1244+026 (hereafter PG1244) and 1H0707-495 (hereafter 1H0707). These
have very similar H$\beta$ line widths, and very similar optical/UV
continua so they form a well matched pair in mass and absolute mass
accretion rate. Both also have high spin as determined from the iron
line/reverberation techniques, though it is clear that this is a much
stronger constraint in 1H0707 (Fabian et al. 2009; Zoghbi et al. 2010;
Kara et al. 2013) than in PG1244 (Kara et al. 2014).  We derive
$L_{bol}/L_{Edd}$ from the outer disc mass accretion rate and find
that both are highly super-Eddington. The disc is unlikely to remain
flat under such conditions, with numerical simulations showing that
such flows should power a strong, clumpy wind from the inner disc
(Takeuchi et al. 2014). Inclination with respect to this wind should
then be an important additional parameter determining the observational
appearance of the inner flow. This is interesting, as while PG1244 and
1H0707 are very similar as regards their optical/UV emission, they
have very different X-ray properties.  1H0707 shows deep X-ray dips,
with extremely strong Fe K$\alpha$ features in these low flux
episodes, whereas the dips (and associated extreme iron features)
are absent in PG1244.  This difference in X-ray behaviour was
classified as `complex' and `simple', respectively (Gallo
2006). Reverberation lags are typically very short in the complex NLS1
(30~s for 1H0707: Fabian et al. 2009) and longer in the simple ones
(200~s for PG1244: Alston, Done \& Vaughan 2014).

We propose that inclination with respect to a strong wind from a
super-Eddington accretion flow is the major underlying difference between
these two different types of NLS1. Absorption in a highly ionised wind
can fit the observed iron features seen in the non-dip spectra of
1H0707, while stochastic occultations by cooler clumps in the wind can
produce the dips and give rise to more complex spectral curvature
(Hagino et al. 2016, see also
Mizumoto, Ebisawa \& Sameshima 2014; Miller et al. 2007; 2010)
These dips also shorten the reverberation time
lags, so that the observed $30~s$ reverberation delay seen in 1H0707
can be produced from an intrinsic lag of 200-300~s, as seen in PG1244
(Gardner \& Done 2015). More face on inclination angles do not
intercept (much of) the wind, so do not show the extreme features at
iron, nor the dips.

This is very different to the standard 'lamppost' model used to derive
the high black hole spin in these objects, where an extremely compact
X-ray source on the spin axis of the black hole illuminates a flat
disc (Miniutti \& Fabian 2004). In this geometry, the dips 
are caused by an extremely compact X-ray source
on the spin axis of the black hole approaching the event horizon. The resulting strong light bending
focusses the intrinsic continuum away from the observer (producing
the drop in flux) so it instead strongly illuminates the very inner
disc, producing a very centrally concentrated emissivity for the line which 
can simultaneously explain the very strongly relativistically smeared, 
reflection dominated emission seen in the dips (Miniutti \& Fabian 2004). 
However, if the lamppost geometry is not
correct, then the spins derived from it need not be correct
either. This could remove the requirement that there are high spin {\em
  radio quiet} objects, re-opening the possibility of the spin-jet
paradigm where high spin is a necessary and sufficient trigger for
highly relativistic jets.

\section{Constraints from relativistic reflection models}

The classic way to measure black hole spin is via the iron line
K$\alpha$ profile in the 2-10 keV bandpass (Fabian et al. 1989; 2000).
1H0707 always shows strong broad features at 6-7 keV, especially 
during the deep
X-ray intensity dips. Fitting reflection
models to these spectra give a spin measurement which is always high
(e.g. Fabian et al. 2004; 2009; 2012). The most complete models of
Zoghbi et al. (2010) give a disk inner radius of $r_{in}=
1.23^{+0.07}_{-0.0}R_g$ (corresponding to spin
$a_*=0.998^{+0.0}_{-0.001}$) for a disc inclination of
$58.5^{+0.8}_{-0.7}$ and emissivity index of $6.6^{+1.9}_{-1.1}$.
While these can all be explained within the lamppost model, the data
additionally require that iron is overabundant by a factor $\sim 9$ in
order to match the observed features.  

Fitting the lag-frequency and lag-energy data from reverberation mapping also gives
constraints on the black hole mass, spin and disc inclination, though there are
significant degeneracies between these parameters from the spectral—timing data alone (e.g. 
Cackett et al. 2014). Assuming that the lag timescale $\Delta t \sim 2H/c = 2hR_g/c$,
where $H$ is the physical height of the lamppost and $h=H/R_g$ is this in terms of gravitational radii,
only constrains the product $h M$ rather than each parameter separately.
In 1H0707, the observed 30~s soft
lag (Fabian et al. 2009; Zoghbi et al. 2010; 2011) corresponds to an intrinsic 80~s light travel time
after correcting for spectral dilution, which 
implies a lamppost height of $4R_g$ for a $2\times 10^6M_\odot$ black
hole (Kara et al. 2013). However, the spectral data discussed above 
strongly require that the lamppost height is small
from the strongly centrally peaked emissivity (e.g. Fabian et al. 2012). Hence for these
data a lamppost height of $2R_g$ implies a black hole mass of $4\times 10^6M_\odot$, though at such low heights
relativistic effects become very important. Light bending will increase the proper path
length, and time dilation (Shapiro delay) will increase the time difference. 
A lamppost source at height $2R_g$ above a flat disc around a maximally spinning black hole has a 
minimum lag time of $2.6-1.235R_g/c$ (face on-edge on) if light travels in straight lines without time dilation. 
Including all relativistic effects lengthens these to around $4R_g/c$ (Dovciak, private communication, see also Dovciak et
al.  2014). Thus a lamppost height of $2R_g$ still implies a $2\times 10^6M_\odot$ black hole.

In sharp contrast, the simple NLS1 PG1244 has no deep intensity dips, 
so the iron features are not so strong and do not of themselves
require high spin (Jin et al. 2013), nor very strong iron overabundance 
(Jin et al. 2013 assume solar, Kara et al. 2013 find iron is $2.1^{+0.4}_{-0.1}\times $
solar). However, interpreting a (rather low significance: see Alston et al. 2014) lag of 1000~s at the iron line energy in a 
reflection dominated scenario gives $r_{in}=1.6^{+0.5}_{-0.3}R_g$ (corresponding to spin $a_*=
0.97^{+0.03}_{-0.29}$) and inclination of $38^\circ\pm 3^\circ$ for a
lamppost height  of $5R_g$ above a $1.3\times 10^7M_\odot$ black hole
(Kara et al. 2014).  

\section{Zeroth order estimates of mass and mass accretion rate from the optical spectrum}

Figure 1 shows the optical spectrum of 1H0707 from CTIO (black,
Leighly \& Moore 2004) together with that of PG1244 from SDSS.  These
are redshift corrected using $z=0.0398$ and $0.0482$, de-reddened with
$E(B-V)=0.073$ and $0.032$ (Bessell 1991; Kalberla et al. 2005), and
fluxed with distances of 168.6 and 206.2 Mpc, respectively.  Note that
the redshift of 1H0707 used here is slightly different to that of
$0.040568\pm 0.000150$ quoted in the NASA/IPAC Extragalactic
Database (NED) from the 6dF survey (Jones et
al. 2009), as this redshift gave a noticeable offset with respect to the FeII lines in PG1244.
The 6dF redshift was calculated from cross-correlation with a
late type galaxy template, using the absorption lines. However,
Leighly \& Moore (2004) detect no absorption lines from the host
galaxy in their spectrum, and instead determined a redshift of $\sim 0.04$ (no uncertainty
given) from the broad MgII and H$\beta$ emission lines. 
We determine our 
redshift by  cross-correlating the
spectrum of 1H0707 with that of PG1244 in the 4000-6700~\AA\ region,
after blacking out the regions where strong narrow lines dominate in
PG1244 but not in 1H0707.  This gives $z=0.03979\pm 0.00084$ factoring
in the uncertainty on the PG1244 redshift from SDSS.  We use this
slightly lower value as we are most interested in a comparison of
1H0707 with PG1244, but the change in implied distance is less than
2\% so this has no significant impact on the luminosity.
We rescale PG1244 by a factor of 0.84 so as to overlay it on top of 
1H0707. Hence the two objects have the same optical luminosity to within 20\%.

Figure 1 shows that 1H0707 has much weaker narrow emission lines than
PG1244, but that the broad emission lines are extremely similar,
especially the strength of the FeII emission line blends and the ratio
of FeII to the broad H$\beta$ (PG1244 has a narrow H$\beta$ line
component which makes it appear stronger). The FeII emission is
clearly not a factor $\sim 5$ larger in 1H0707 than in PG1244, so this
does not support the difference in iron abundance suggested by the
relativistic reflection fits (Section 2), though we note that the
FeII strength depends also on $L/L_{Edd}$ as well as abundance (Ferland et al. 2009).
This is discussed further in Section 5.2. 

\begin{figure}
\epsfxsize=0.47\textwidth \epsfbox{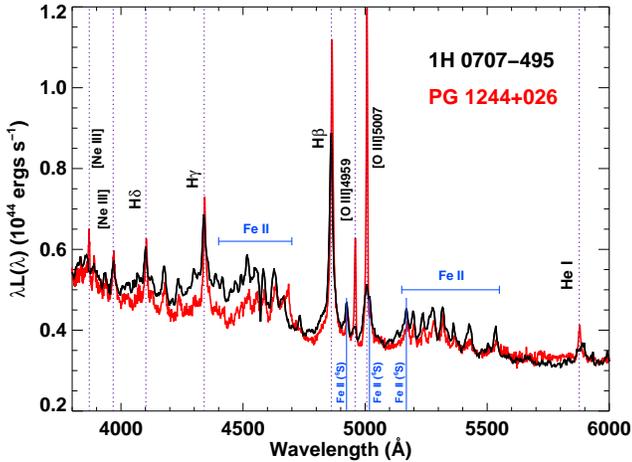}
\caption{The optical spectra from 1H0707 (black: Leighly \& Moore 2004) and
  PG1244 (SDSS, red, rescaled by 0.84 to match 1H0707).  The overall
  luminosities are similar, as are the lines (e.g. FeII and broad lines), except
  that 1H0707 has much weaker forbidden lines and narrow lines (e.g. [OIII]).}
\label{fig:optical}
\end{figure}

The H$\beta$ line width is determined as in Jin et al. (2012a),
i.e. after subtraction of the strong Fe II lines, and subtraction of a
narrow component of H$\beta$ with width fixed to that of the [OIII]
lines.  This gives very similar H$\beta$ FWHM of 980 and 940 km/s, and
$\lambda L_{5100}=3.3$ and $3.9\times 10^{43}$~ergs~s$^{-1}$ 
for 1H0707 and PG1244,
respectively.  Thus they have very similar mass estimates from H$\beta$, with $
M\sim 4\times 10^6M_\odot$ or 
$M\sim 2\times 10^6M_\odot$ depending on whether the scaling
relations of Vestergaard \& Peterson (2006) or Woo \& Urry (2002) are
used. The similarity in line widths do not at all support the factor of nearly 10 difference
in mass from the reverberation results. 

The bolometric luminosity is typically estimated as $9\times \lambda
L_{5100}$ (e.g. Kaspi et al. 2000), so both these sources have
$L_{bol}\sim 3\times 10^{44}$~ergs~cm$^2$~s$^{-1}$.  This is close to
the Eddington limit, $L_{Edd}=2.6(5.2)\times
10^{44}$~ergs~s$^{-1}$ for a black hole mass of $2(4)\times
10^6M_\odot$. However, bolometric correction
$\kappa_{5100}=L_{bol}/\lambda L_{5100}$ clearly changes with mass and
mass accretion rate. Several studies have shown that the `standard'
value of $9\lambda L_{5100}$ used above is only appropriate for the
broad line Seyferts ($L/L_{Edd}~<~0.2$), while objects around
$L_{Edd}$ require correction factors of 20-40 (Vasudevan \& Fabian
2007; 2009, Jin et al. 2012a; b).  This would put both 1H0707 and
PG1244 at $L_{bol}\sim 10^{45}$~ergs~s$^{-1}$, strongly
super-Eddington for either black hole mass. 

At such high Eddington fractions it is clearly debatable as
to whether the H$\beta$ scaling relations hold. Marconi et al. (2008)
suggest a correction to the mass scalings such that the BLR clouds
trace the effective gravity.  Using this, the black hole masses
increase to around $10^{7}M_\odot$, with  $L_{Edd}=1.3\times
10^{45}$~ergs~s$^{-1}$, so the accretion flow is still close to 
the Eddington limit as required for the higher bolometic correction. 

The radiation pressure correction to the mass is not widely used as
the general lack of observed wind features for the low ionisation BLR
lines form a strong argument against such effects being important
(Baskin, Laor \& Stern 2014a). Physically, this could indicate that
the clouds are optically thick to electron scattering, with columns of
$>10^{24}$~cm$^{-2}$ (consistent with the low ionisation broad lines
forming in the disc: Collin-Souffrin et al. 1980), or that the
radiative acceleration only affects the front face of the cloud
(Baskin, Laor \& Stern 2014b). However, part of the H$\beta$ line is
clearly blue-shifted (Figure 1), indicating that these objects could
be so extreme that even part of the low ionisation lines (as well as
the high ionisation lines such as CIV) come predominantly from a wind
(see also Leighly \& Moore 2004; Leighly 2004).

The new results from reverberation mapping (see Section 2) 
strongly require that the
radiation pressure corrected mass is not appropriate, and that the
mass for 1H0707 is  $\le 4\times 10^6M_\odot$. The similarity of the 
H$\beta$ emission line widths support a similarly low mass for PG1244. 

\begin{figure*}
\begin{tabular}{cc}
\epsfxsize=0.47\textwidth \epsfbox{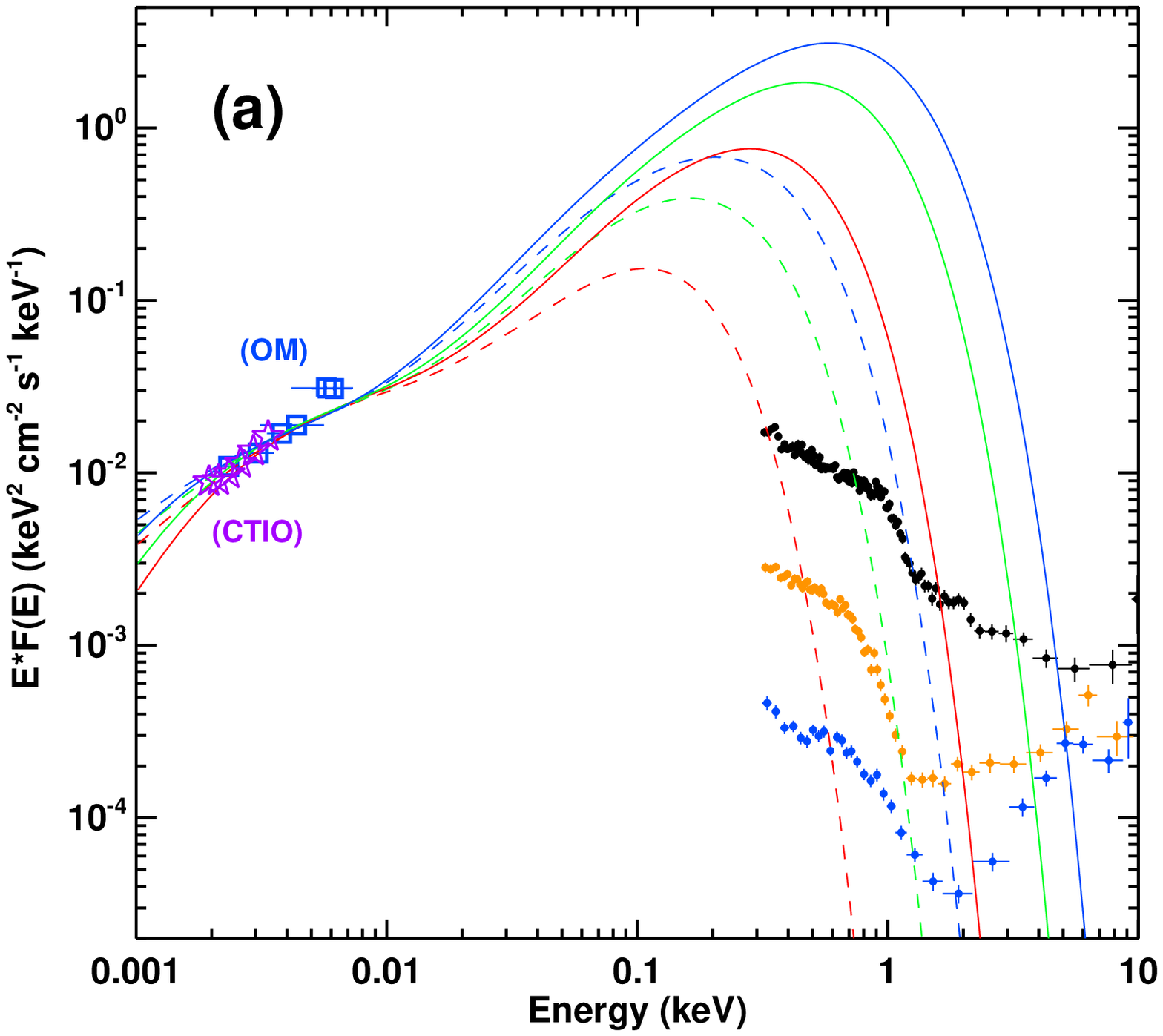} &
\epsfxsize=0.47\textwidth \epsfbox{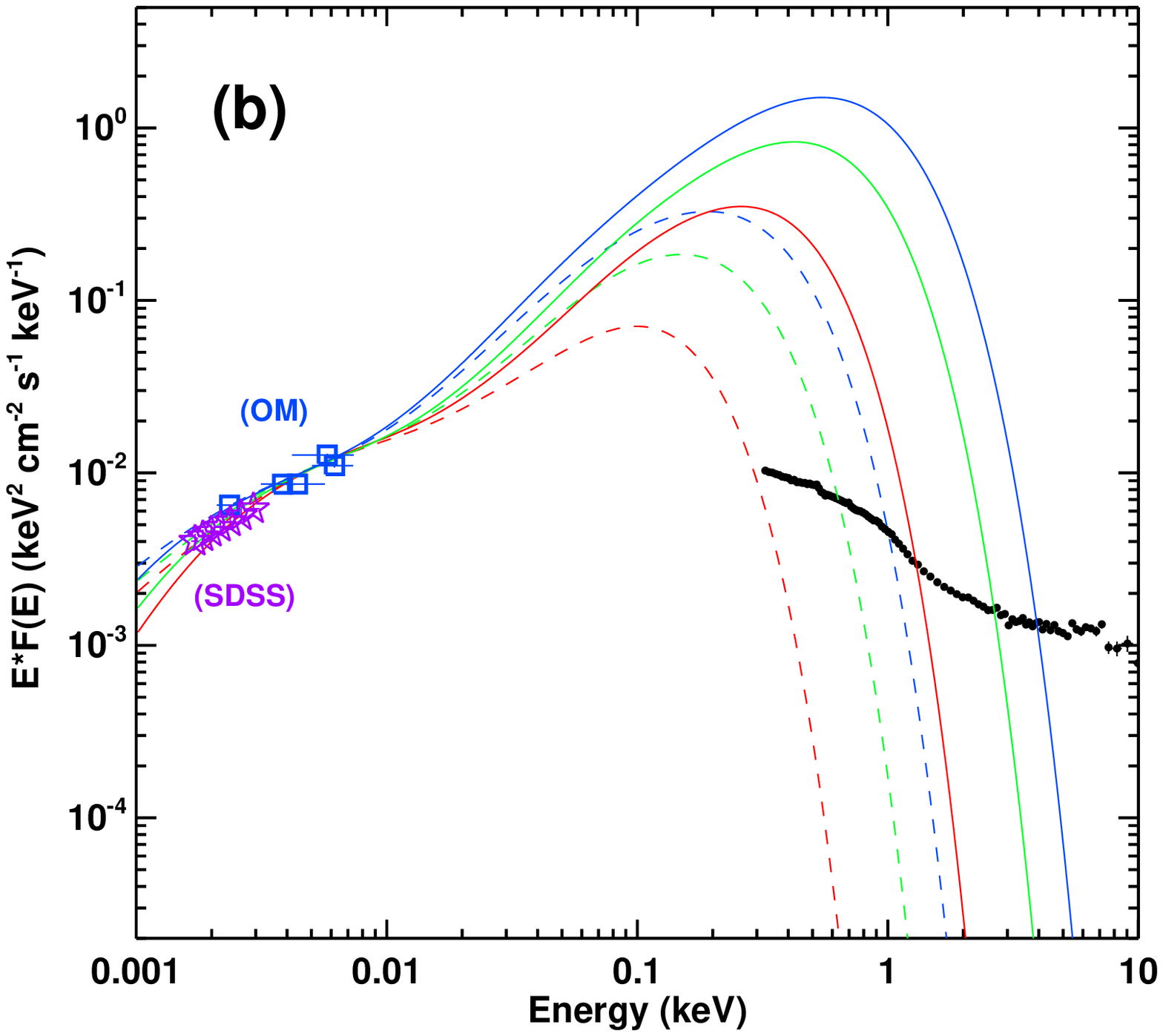}\\
\end{tabular}
\caption{Continuum fitting with {\sc optxconv} to the NLS1 a) 1H 0707-495 and b)
  PG 1244+026. The red, green and blue solid lines show the disc emission
  inferred for an inclination of $30^\circ$ from the optical/UV continuum for
  spin of $a=0,0.9$ and $0.998$ respectively for a mass of $2\times
  10^6M_\odot$, while the dashed lines show the same for a mass of
  $10^7M_\odot$. For 1H 0707-495, the optical continual points (star symbol)
  extracted from the CTIO spectrum were scaled up by 50\% to account for the
  discrepancy with OM data due to long term variability and/or aperture
  difference.}
\label{fig:sed}
\end{figure*}

\section{Broadband spectra: mass accretion rate and inclination}

We assemble simultaneous optical/UV/X-ray spectral energy
distributions (SED) from XMM-Newton data, using the OM to give optical
and UV photometric points along with the X-ray data from the EPIC pn
camera. All XMM-Newton data reduction and spectra extraction were
performed following the the standard procedures with {\tt SAS} v14.0.0
and the latest calibration files.  For both 1H0707 and PG1244, we
extend the SED down to lower energies using continuum points from the
optical spectra discussed above. While these data are not
simultaneous, the spectra match extremely well for PG1244, whereas for
1H0707 the optical spectrum discussed in Section 2 is 40\% lower than
the OM data. This is probably due to long term variability, but the
CTIO spectrum was not fluxed (Leighly, private communication), so part
of this may also be due to slit losses. 

NLS1 in general are less optically variable than
Broad Line Seyferts (e.g. Ai et al. 2013, but see Kelly et al. 2013),
despite having higher X-ray variability (e.g. Leighly et al. 1999;
Ponti et al. 2012).  1H0707 has no optical monitoring data, but UV
monitoring by the XMM-Newton OM (UVW1 filter) shows factor 30\%
variability on timescales of years but less than 2\% variability within
a single observation of $\le$ 100 ks timescale
(Robertson et al. 2015; see also Smith \& Vaughan
2007 for similar limits on fast variability in UVW2).

By contrast, 1H0707 shows dramatic X-ray variability. The source flux
can drop by a factor of $\sim 30$, accompanied by strong variation in
spectral shape (e.g. Zoghbi et al. 2010).  We systematically searched
all available XMM-Newton observations for data which spanned the range
in X-ray shape and intensity. We extracted pn spectra from every 10 ks
segment of each observation, and selected the highest spectrum (Figure
2a: black spectrum from the 60-70 ks segment in the observation on
2010-09-15, OBSID: 0653510401). The lowest spectrum has a low count
rate, so we combine two very similar spectra from 40-50 ks and 50-60
ks segments in the observation on 2011-01-12 (OBSID: 0554710801) to
obtain a low state spectrum with better S/N (Figure 2a: blue
spectrum).  We also select an intermediate state (Figure 2a: orange
spectrum) from the 10-20 ks segment in the observation on 2007-05-14
(OBSID: 0506200301).  The OM data for these observations are
remarkably constant.  The largest difference is 10\% in the UVW2 flux
between 2007-05-14 (intermediate-state X-ray spectrum) and 2010-09-15
(high-state X-ray spectrum), the other differences in the same OM
filters are all less than 4\%. Since only the 2010-09-15 observation
has exposures in six OM filters (i.e. UVW2, UVM2, UVW1, U, B, V), we
used this dataset in Figure 2a. A similar range in X-ray spectra is
shown by Vasudevan et al. (2011), and they also find that this is not
accompanied by strong optical/UV variability.

For PG1244, a 6 month optical reverberation mapping campaign failed to
constrain the black hole mass as there was no significant 
variability (K. Denny, private
communication). We extracted the two archival IUE observations of
PG1244, taken 10 years apart, and these show variability by a factor
of $\sim 2$, similar to the variability seen between the OM and GALEX
UV photometry (Jin, Done \& Ward 2013, hereafter: J13).
Similarly, the 2-10~keV band in PG1244 varies by
a factor 2 on both long and short timescales (J13 and the compilation
of Ponti et al. 2012). The spectral shape remains more or less
constant during this variability, so we use pn data from the longest
XMM-Newton observation (123 ks XMM-Newton on 2011-12-25, OBSID:
0675320101) to define the typical source spectrum (J13; Done et al. 2013, hereafter: D13).
We use the simultaneous OM data from this observation which matches well to
the SDSS spectrum.

Figures 2a and b show the resulting SEDs for 1H0707 and PG1244,
respectively, where we de-absorb the X-rays using only the Galactic
gas column of $N_H=4.31$ and $1.87\times 10^{20}$~cm$^{-2}$ (Kalberla
et al. 2005), and de-redden the corresponding optical/UV values of
$E(B-V)=0.073$ and $0.032$ (Bessell 1991). The resultant SEDs are then
de-redshifted to the AGN rest frame. These are the lowest possible
absorption values as they only include material in our galaxy. Any
additional material in the host galaxy will 
increase the intrinsic UV/soft X-ray flux, and hence increase the
inferred mass accretion rate. 

\begin{table*}
\caption{Comparison of the {\sc optxconv} results including the self
consistent colour temperature correction with those from {\sc kerrbb}
with $f_{col}=1$. Both disc models give similar (within a factor $\sim
2$ in $\dot{m}$ from the optical/UV continuum fits as both have
$f_{col}=1$ in the optical region. The X-ray excess is defined as the
disc model flux in the 0.3-2~keV bandpass, 
absorbed by the Galactic column, divided by the observed flux. 
of $1.39\times 10^{-11}$~ergs~cm$^{-2}$~s$^{-1}$ (the maximum) for 1H0707, 
and $1.33\times 10^{-11}$~ergs~cm$^{-2}$~s$^{-1}$ for PG1244. 
The {\sc optxconv} models
extend further into the soft X-ray bandpass due to their
$f_{col}\sim 2$ in the EUV region, so require larger losses. 
}
\begin{tabular}{l|l|l|l|l|l|l|l}
 \hline
   \small source & M & i & spin & $\dot{m}$(optx) 
& X-ray excess(optx) &  $\dot{m}$(kerrbb) & X-ray excess(kerrbb)\\
\hline
1H0707 & $2\times 10^{6}$ & 30 & 0 & 20 & 36 & 14 & 2.1 \\
1H0707 & $2\times 10^{6}$ & 30 & 0.9 & 59 & 162 & 35 & 27 \\
1H0707 & $2\times 10^{6}$ & 30 & 0.998 & 146 & 360 & 63 & 61 \\
\hline
1H0707 & $10^{7}$ & 30 & 0 & 0.9 & 0.24 & 0.6 & 0.0 \\
1H0707 & $10^{7}$ & 30 & 0.9 & 2.5 & 4.9 & 1.5 & 0.02  \\
1H0707 & $10^{7}$ & 30 & 0.998 & 6.3 & 19 & 2.7 & 0.20\\
\hline
1H0707 & $2\times 10^{6}$ & 60 & 0 & 44 & 90  & 31 &  11\\
1H0707 & $2\times 10^{6}$ & 60 & 0.9 & 121 & 390 & 79 & 140 \\
1H0707 & $2\times 10^{6}$ & 60 & 0.998 & 265 & 755 &142 &  350\\
\hline
1H0707 & $10^{7}$ & 60 & 0 & 1.9 & 1.9 & 1.4 & 0.002 \\
1H0707 & $10^{7}$ & 60 & 0.9 & 5.1 & 28 & 3.4 & 1.5  \\
1H0707 & $10^{7}$ & 60 & 0.998 & 11 & 92 & 6.0 & 10 \\
\hline
PG1244 & $2\times 10^{6}$ & 30 & 0 & 13 & 18 & 9.3 & 1.0 \\
PG1244 & $2\times 10^{6}$ & 30 & 0.9 & 38 & 80 & 23 & 14\\
PG1244 & $2\times 10^{6}$ & 30 & 0.998 & 94 & 175 & 42 & 31\\
\hline
PG1244 & $10^{7}$ & 30 & 0 & 0.6 & 0.1 & 0.4 & 0.0 \\
PG1244 & $10^{7}$ & 30 & 0.9 & 1.7 & 2.3 & 1.0 & 0.01 \\
PG1244 & $10^{7}$ & 30 & 0.998 & 4.1 & 9.0 & 1.8 & 0.07 \\
\hline
\end{tabular}
\label{tab:lledd}
\end{table*}

The optical/UV emission looks like a disc spectrum in both objects, 
with no obvious flattening at red wavelengths 
which would indicate substantial host galaxy contamination (see e.g. the
spectrum of IRAS 13224-3809 in Figure 1 of Leighly \& Moore 2004). Similarly,
there are no obvious absorption lines from the host galaxy in the spectra of either
object, again, indicating that any host galaxy contamination is low. 
Hence 
we fit with a disc model to estimate the mass accretion rate. We use
the {\sc optxconv} model of D13, as this approximates the full
radiative transfer, fully relativistic models of Hubeny et al. (2001)
(updated in Davis, Woo \& Blaes 2007), but is more flexible in
spectral fitting as it is analytic rather than limited to precomputed
tables. The intrinsic disc spectrum is calculated using a colour
temperature correction factor, $f_{col}$, to approximate the effect of
electron scattering in the disc photosphere.  This colour temperature
correction factor is not a constant in these models, but is dependent
on the photosphere temperature. At small radii, the effective
temperature is in the far UV/soft X-ray, where the true absorption
opacity is very low. The photosphere is set by the point at which the
effective optical depth, $\tau_{eff}\approx
\sqrt{\tau_{es}\tau_{abs}}$ is unity, where $\tau_{es}, \tau_{abs}$
are the electron scattering and true absorption optical depth,
respectively.  Photons can escape after multiple scatterings from deep
inside the disc where the temperature is higher giving rise to a
spectrum which is approximately $B_\nu(f_{col}T)/f_{col}^4$ with
$f_{col}\sim 2$. Conversely, in the outer regions of the disc, where
the temperature is in the optical, the true absorption opacity is high
so $f_{col}\sim 1$. This change in colour temperature between the
optical and far UV predicts a change in curvature of the disc spectrum
in the UV (Done et al. 2012), and this intrinsic spectrum
then has relativistic effects superimposed to give the observed
emission (D13).

At a fixed energy, $h\nu~<<~kT_{peak}$ where $T_{peak}$ is the
temperature at which the disc luminosity peaks, a standard
Shakura-Sunyaev disc has $L_\nu\propto (M\dot{M})^{2/3}\cos i$
(e.g. Davis \& Laor 2011). This uses only the 
optical/UV flux to determine the
mass accretion rate through the outer disc. However, standard discs
have constant mass accretion rate with radius, so this also sets the total 
$L_{bol}=\eta\dot{M}c^2$. Hence $\dot{m}=\dot{M}/\dot{M}_{Edd}\propto
\eta L_\nu^{3/2} /(M^2\cos^{3/2} i)$ where $\eta\dot{M}_{Edd} c^2
=L_{Edd}\propto M$ and the efficiency $\eta=0.0572, 0.155$ and $0.321$
for $a_*=0,0.9 $ and $0.998$ as the innermost stable circular orbit
decreases from $r_{isco}=6, 2.32$ to $1.236$.  Hence the observed
optical/UV continuum determines $\dot{m}$ from the combination of
$\eta /(M^2\cos^{3/2} i)$, and this is maximised for the smallest
mass, highest spin, highest inclination. The analytic approximation gives a nice way to
see the effect of all the parameters, and indeed in the figures below 
it is clear that even the further UV datapoint is still on the Rayleigh-Jeans tail
of the disc spectrum so that the approximations should be valid, 
but we stress that we use full spectral 
fitting to derive all mass accretion rates given below. 

For 1H0707 solid red, green and blue lines in Figure 2a show the results
assuming a black hole mass of $2\times 10^6M_\odot$, for a spin of 0
(red), 0.9 (green) and 0.998 (blue) assuming a standard AGN
inclination of $30^\circ$.  The derived mass accretion rate through the
outer disc is $\dot{m}=20, 63$ and $151$, respectively, with
increasing luminosity and peak temperature with black hole spin as
expected from the discussion above. Even for the higher mass of
$10^7M_\odot$, derived from applying the radiation pressure correction
to the H$\beta$ line width, gives $\dot{m}=0.9, 2.7$ and $6.6$,
respectively, though the bolometric luminosity and peak temperature
are now both lower by a factor of $\sim 5$ and $3$, respectively for
each spin. 

Figure 2b shows the same sequence of models for PG1244 for 
$i=30^\circ$. The derived mass accretion rate is slightly lower, with 
$\dot{m}=14, 43$ and $110$ for a mass of $2\times 10^6M_\odot$, 
compared to $\dot{m}=0.6,1.9$ and $4.8$ for $10^7M_\odot$.

For 1H0707 we also explore the effect of a higher inclination, as
required by the relativistic reflection fits (Section 2).  This gives
$\dot{m}=44, 120$ and $269$ for the mass of $2\times 10^6$ or
$\dot{m}=1.9, 5.0$ and $11$ for $10^7M_\odot$ for an inclination of
$60^\circ$. The fits are all summerized in Table 1. It is clear that
only the very highest mass, lowest spin and lowest inclination
solutions have sub-Eddington accretion flows.  This is exactly the
opposite requirement to the 
relativistic reflection fits in 1H0707, which need
low mass, high spin and high inclination (Section 2). 

This conclusion does not strongly depend on the colour temperature
correction in {\sc optxconv} as this is close to unity in the optical,
reaching a maximum of around 1.2 for the highest temperatures which
contribute to the UV data used here. This decreases the UV
flux slightly, as it pushes it to higher energies (see e.g. Figure 1a in
Done et al. 2012). Thus the observed UV flux can be fit with a slightly
lower mass accretion rate if there is no colour temperature
correction. However, this is in general a rather small effect.  We
demonstrate this explicitly by repeating the fits with the {\sc
  kerrbb} disk model, fixing $f_{col}=1$. The inferred mass accretion
rates are around a factor 2 smaller for the same mass/spin, but all
the fits with high spin are still highly super-Eddington (see Table
1).

However, the colour temperature correction does have a large effect on
the predicted soft X-ray extent of the disc emission as the peak disc
temperature is boosted by a factor $\sim 2$ by this effect (see
e.g. Figure 1a in Done et al. 2012).  
The peak disc
temperature is $T_{peak}\propto f_{col} [(\dot{M}
M)/R^3_{isco}]^{1/4} \propto f_{col}[ (L_\nu/\cos i)^{3/2}/(M^3 r^3_{isco})]^{1/4}$
so it is highest for the lowest mass, highest spin, highest
inclination solutions with highest colour temperature. 
Figure \ref{fig:sed}a and b show the high energy extrapolation of the {\sc
  optxconv} model SED fits to the optical/UV. It is clear that all
these dramatically over-predict the soft X-ray flux except for for the
lowest spin, highest mass black hole models. We quantify this by
integrating the disc model flux in the 0.3-2~keV bandpass, including
the effect of Galactic absorption, and divide this by the observed
0.3-2~keV flux. For PG1244 we use the average flux of 0.3-2~keV flux
of $1.33\times 10^{-11}$~ergs~cm$^{−2}$~s$^{-1}$, while for 1H0707 we
pick the maximum observed flux of $1.39\times
10^{-11}$~ergs~cm$^{-2}$~s$^{-1}$. These ratios are typically much
larger than unity for all except the highest mass, lowest spin, lowest
inclination models for the colour temperature corrected models, while
they are substantially reduced by removing the colour temperature
correction factor (see Table 1). However, even without a colour
temperature correction, all high spin, high inclination solutions for
1H0707 all dramatically over-predict the soft X-ray flux.

\section{Discussion}

1H0707 is the most convincing candidate for extremely high black hole
spin as derived from the reflection/reverberation fits to the X-ray
spectral-timing properties. However, these fits require a low mass,
high spin black hole, viewed at high inclination, i.e. exactly the case
where the optical/UV fits imply the most strongly super-Eddington
flows. This means that the inner disc is highly unlikely to
be flat. Since a flat disc was assumed in the lamppost geometry to
derive these constraints on mass, spin and inclination, then clearly
these need not be robust. We critically examine these below.

\subsection{Alternative models for the X-ray spectra and variability
  of 1H0707?}

\begin{figure}
\begin{tabular}{c}
\epsfxsize=0.47\textwidth \epsfbox{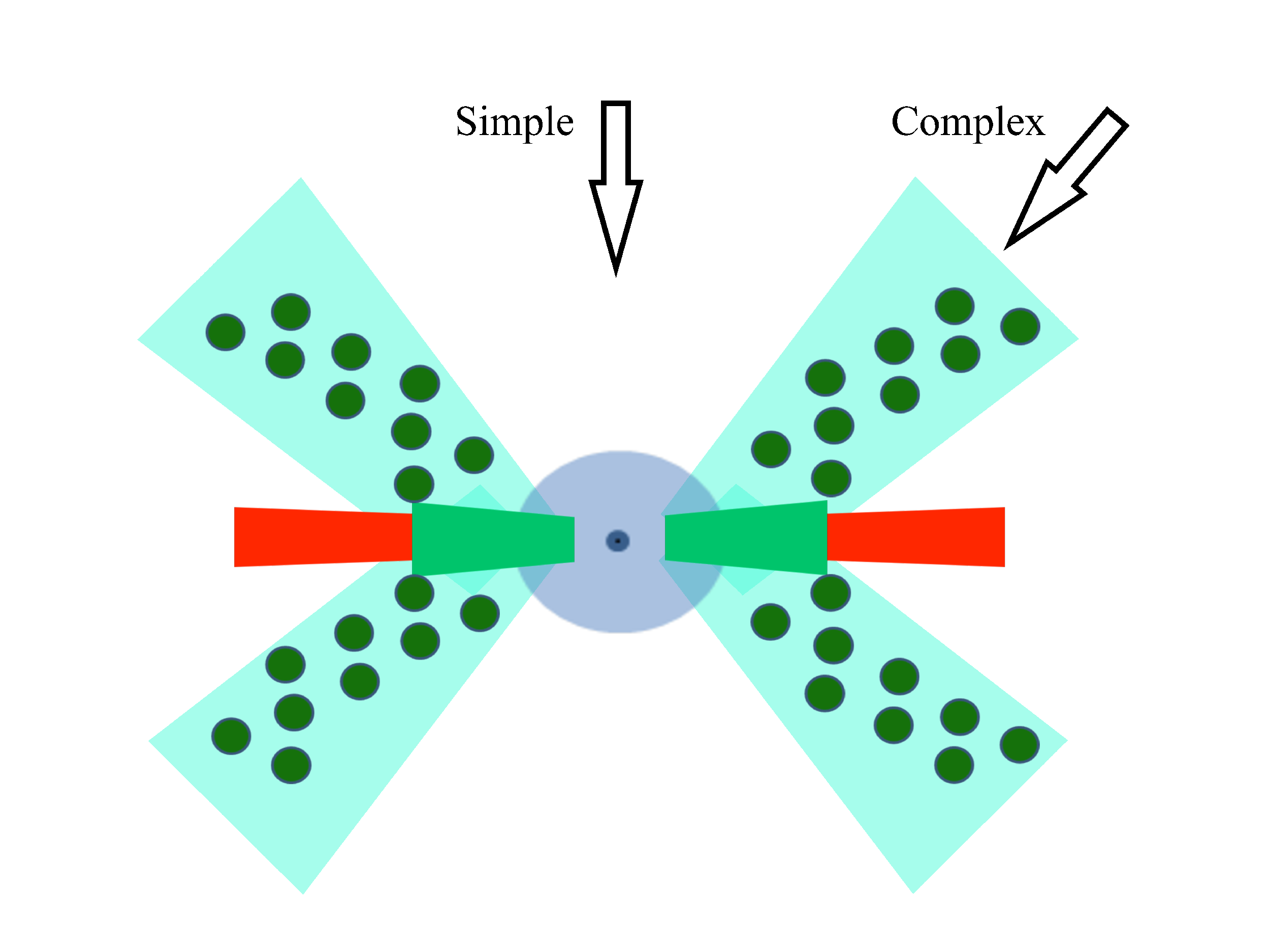} \\
\end{tabular}
\caption{A schematic picture of the clumpy wind geometry from a super-Eddington flow 
which could explain the difference in  X-ray properties between simple and complex NLS1. The hard X-ray
corona (inner blue flow) is absorbed by the wind at high inclinations, with the smooth, highly ionised 
absorption (cyan) producing a strong drop at the (blue-shifted) line energy, while  
cooler, less ionised clumps (green), produce stochastic dips and complex absorption at all energies (Hagino et al. 2016).}
\label{fig:sketch}
\end{figure}

The dramatic X-ray spectral features in 1H0707 are so broad and strong
around iron that probably all relativistic reflection models will
require high spin, irrespective of the specific reflector geometry. However, it
is possible to fit the spectral features with a very different model,
based on absorption in a clumpy wind (Hagino et al. 2016). This is
shown schematically in Figure \ref{fig:sketch}. 
Winds are generically expected to be clumpy, and the super-Eddington
wind simulations of Takeuchi et al. (2014) show clumps forming from 
the (almost) standard Rayleigh-Taylor instability. However, the
material is also illuminated by X-rays and in pressure balance, so
there is an ionization instability which produces
two distinct phases in the gas, resulting in clumps of cooler, less
ionized material embedded in less dense, highly ionised gas (Krolik,
McJee \& Tarter 1981). 

Hagino et al. (2016) model the hot phase of the wind, and include cool
clumps in a phenomenological way to show that
lines of sight which
intercept the wind always show deep absorption features at iron from
the hot, highly ionised phase of the wind, together with stochastic
dips and complex absorption when the less ionized clumps occult the
X-ray source. This model can give as good a fit as relativistic
reflection to all the 2-30~keV
data (XMM-Newton and NuSTAR) from 1H0707, and does not require a high
spin black hole (Hagino et al. 2016). This is rather different to the 
single ionisation wind model, which still requires a contribution from
highly relativistically smeared reflection 
(Dauser et al. 2012).

The clumpy wind absorption model also has the potential to explain the short
reverberation lags observed.  Gardner \& Done (2015) show how a low
spin, full spectral-timing model designed to explain the 200~s soft lag seen in
PG1244 (Gardner \& Done 2014a) can also match the lag-frequency data
from 1H0707 just by including occultations. 
In their non-relativistic analysis, the occultation timescale is simply the time taken for the clump 
to cover the source. This is $\Delta t\sim xR_g \sqrt{r} /c$ for a 
a source of size $xR_g$, covered by a cloud on a circular orbit at $rR_g$ 
of size scale larger than the source. The deep dips in 1H0707 take
$\sim 200$s (Fabian et al. 2012), 
so this would imply $x\sqrt{r}=20$ or clumps orbiting at $r=11$ for a source size of $6R_g$
and a black hole mass of $2\times 10^6M_\odot$. It remains to be seen whether the clumpy winds
which are generically produced from super Eddington flows (e.g. Takeuchi et al. 2014) can match these
properties.

Similarly, it remains to be seen whether occultations can match the spectral behaviour, 
both during the dips (time resolved spectra) and the high frequency 
lag-energy (Fourier resolved) spectra. The rather simple model of occulting clouds in 
Gardner \& Done (2015) did not give a good fit to the 
dip spectra from 1H0707, but this model did not include 
reflection from the complex material on the far side of the source,
and their clumps were highly ionized and single phase, rather than low
ionization material embedded in a more highly ionized wind (see the sketch 
of the proposed geometry in Fig.\ref{fig:sketch}). The high frequency lag-energy spectra 
also contain features around the iron line energy, and again these require a 
simulation of the lag-energy spectrum expected from the full geometry of Fig.\ref{fig:sketch}
in order to see whether this can match the data. We note that the lamppost model itself 
cannot easily match the high frequency lag-energy spectra seen in 1H0707
(Kara et al. 2013; Wilkins et al. 2016), showing that the source geometry is almost certainly more complex. 

One objection to the idea of cool clumps providing some of the
variability in 1H0707 is that they will produce an iron fluorescence
line if the clumps subtend a reasonable solid angle to the X-ray
source.  The line is smeared by the velocity
structure of the wind, but without extreme relativistic effects it
should be observable as a distinct feature rather than broadened into
the red wing (Zoghbi et al. 2010). However, if the clumps are embedded
in a more highly ionized wind, then Compton scattering in the highly
ionized phase will reduce the observed line flux. Proper simulations
of this geometry are clearly required in order to test whether 
the predicted line can indeed match the observed spectra from 1H0707. 

\subsection{The difference between simple and complex NLS1}

The wind model sketched above gives a potential geometric explanation
for the large difference in X-ray spectral behaviour between simple
and complex NLS1.  This difference must have its origin in a
difference in one of the parameters which can affect the appearance of
the black hole accretion flow, but several sources are seen to
transition between simple and complex behaviour (Gallo 2006).
Thus the difference cannot originate with a fixed
parameter such as spin. Instead, it must be associated with something
which can vary. Figure \ref{fig:sketch} clearly shows that inclination
angle with respect to the wind can make a large difference in the
observed properties of the inner X-ray source. Numerical simulations show that
the polar and opening angle of the wind can fluctuate stochastically,
as well as vary with $L/L_{Edd}$, so the inclination angle with
respect to the wind could be the required parameter determining the
difference between simple and complex NLS1. 

1H0707 (complex NLS1) and PG1244 (simple NLS1) have very similar
H$\beta$ line widths and optical luminosity, so have similar mass and
mass accretion rates to zeroth order, and their masses are most
probably low just from X-ray variability timescale arguments. If
1H0707 is at higher inclination than PG1244 then it should have higher
$L/L_{Edd}$ for the same black hole spin. We take as an example 1H0707
at $60^\circ$ and PG1244 at $30^\circ$, both with low spin and mass of
$2\times 10^6M_\odot$. The colour(non-colour) temperature corrected
discs give a mass accretion rates of 44(31)$\times$ Eddington for
1H0707 and 13(9) for PG1244, so both are still highly super-Eddington,
but 1H0707 is more extreme that PG1244, as required by the optical
spectra shown in Figure 1, where 1H0707 lacks the narrow [OIII]
emission lines. Weak narrow lines are a characteristic of low
ionisation Broad Absorption Line (BAL) QSO's (Boroson \& Meyers 1992),
where this is likely due to shielding of the narrow line region by
obscuration in a wind (see also Leighly 2004).

If 1H0707 is accreting at a higher Eddington ratio, this should also
affect other features of the spectrum.  $L/L_{Edd}$ is known to
affect the ratio of equivalent widths of emission lines in the broad line region,
particularly FeII/(broad) H$\beta$
(Ferland et al. 2009). The iron blends shown for both NLS1 in Figure 1
are quite similar, and similar also in terms of their ratio of
FeII/broad H$\beta$. If the correlation continues at these high
$L/L_{Edd}$ then 1H0707 should have a slightly lower iron abundance
that PG1244 in order to produce the same ratio of FeII/broad H$\beta$.
However, it seems likely that the FeII/H$\beta$ ratio would also be
affected by geometry of the wind, so the stronger wind in a more
highly super-Eddington flow in 1H0707 could compensate for this.

Eddington ratio is also known to correlate strongly with the 2-10~keV
spectral index (Shemmer et al. 2006; Vasudevan \& Fabian 2007; 2009;
Jin et al. 2012c). 1H0707 should then have a higher $\Gamma$ than
PG1244, though this is poorly constrained from spectral fits.  Simple
wind models fit to 1H0707 give $\Gamma=3.5\pm 0.8$ (Done et al. 2007),
while more complex wind models are consistent with $\Gamma=2.6$
(Hagino et al.  2016), whereas PG1244 has $\Gamma\sim 2.5$
(J13). Hence the data are consistent with a steeper index in 1H0707 as
expected if it does have a higher $L/L_{Edd}$, but do not
significantly require this.

Thus there is no obvious inconsistency with a model for these two NLS1
where they have similar low masses and spins but where 1H0707 has
higher inclination, and hence its similar optical flux requires
higher mass accretion rate. Both are super-Eddington, but 1H0707 is
more super-Eddington that PG1244, so has a stronger wind which
explains its lack of narrow optical line emission by shielding the
NLR, suppressing the narrow [OIII] lines compared to PG1244 (Leighly
2004).  The combination of more highly super-Eddington flow leading to 
stronger mass loss, together with higher inclination
angle means that the wind is more likely to be in the line of sight
for 1H0707.

This interpretation has much less soft X-ray emission than the low
mass, high spin models required by the relativistic reflection
interpretation. Nonetheless, the observed soft X-ray flux is still a
factor $90(11)\times$ larger than the maximum observed from 1H0707,
and $18(1.0)\times$ that observed in PG1244. Thus there is still a
strong requirement of energy loss from the inner disc in 1H0707 and
probably in PG1244 also.  Strong energy losses are easily accommodated
within the super-Eddington flows, as they can advect some of the
power, and/or use it to launch a wind. It could also potentially
launch an accretion powered jet, but both 1H0707 and PG1244 are very
radio quiet, so substantial jet losses seem unlikely.  A similarly
extreme SED shape, with large advection/wind losses is seen in one of
the lowest mass AGN in the local Universe, GH08 (aka RX~J1140.1+0307:
Miniutti et al. 2009; Jin, Done \& Ward 2016). A low spin solution
requires less extreme energy loss, but this is still over-predicts the
observed 0.3-2~keV luminosity by a factor $\ge 11$ for the low
mass/high inclination solution for 1H0707, even with no colour
temperature correction.

\subsection{Spin and the trigger for highly relativistic jets}

Highly relativistic jets are the most dramatic pointer to an accreting
black hole. Apparently superluminal radio blobs can be explained only
if the bulk Lorentz factor, $\Gamma$, is high, and if the angle,
$\theta$, between this motion and the observer's line of sight is
small. The intrinsic emission of the blob is blue-shifted by a factor
$\delta$ and brightened by a factor $\sim \delta^4$, so the jet
emission from the core appears much more luminous in aligned versus
non-aligned objects.  This means that they are extremely bright in
radio (synchrotron from the jet) and also in high energy GeV
gamma-rays (synchrotron self Compton or External Compton from the jet:
e.g. Ghisellini et al. 2010).

Another pointer to such relativistic jets is the radio lobe emission
on much larger (kpc) scales. This is the interaction of the jet with
its (stationary) environment, so is easily visible in non-aligned
objects. Population studies show that the Fanaroff-Riely type I and II
galaxies are consistent with being the misaligned versions of the
highly beamed BL Lacs and Flat Spectrum Radio Quasars (FSRQ), respectively
(Urry \& Padovani 1995). 

Most AGN do not show such large scale radio structures, but weak
core radio emission is ubiquitous. Clearly, the radio luminosity can
also be dependent on the accretion power, so a better way to
parameterise the importance of the radio jet is via the radio-loudness
parameter, $\cal R$, defined as the ratio of flux at 5 GHz to that in
the optical B band. This ratio changes by more than a factor 1000,
even when restricted to black holes of the same mass (Mclure \& Jarvis
2004), or to black holes of the same $L/L_{Edd}$ (Sikora, Stawasz \&
Lasota 2007). Clearly, it is very important to understand the origin
of this spread in radio power, not only to understand relativistic
jets but also to track their impact on their host galaxy via jet mode
feedback.

Black hole binaries give some insight into the origin of this spread.
These show a dramatic transition in their accretion flow properties at
Eddington scaled mass accretion rates $L/L_{Edd}\le 0.01$. This most
probably marks the transition from a geometrically thin disc at high
mass accretion rates (Shakura \& Sunyaev 1973) to a radiatively
inefficient flow (ADAF: Narayan \& Yi 1995; Esin et al. 1997), see e.g.
the review by Done, Gierlinski \& Kubota (2007). This change in
accretion mode is strongly correlated with radio jet properties in the
BHB, with a steady, compact jet seen at low luminosities, which
collapses, often by ejection of discrete blobs, when the flow makes a
transition to a thin disk (Fender, Belloni \& Gallo 2004). Thus a
single black hole binary can make a transition from being radio loud
to radio quiet (Maccarone, Gallo \& Fender 2003). However, none of
this radio emission is ever associated with a highly relativistic
outflow. Even the ejected blobs have speeds which are only mildly
relativistic, at $\Gamma< 2$ (Gallo, Fender \& Pooley 2003) while the
similarities of the radio-X-ray correlation seen in binaries with
very different inclinations shows that the steady compact jet has
$\Gamma\le 1.7$ (Gallo, Fender \& Pooley 2003).
Similarly, the ubiquitous radio emission in
radio-quiet AGN is also not from a highly relativistic jet, as it has
comparable in both high and low inclination sources (Seyfert 2 and
Seyfert 1's: Thean et al. 2001).

Thus the radio emission seen in `radio-quiet' AGN is mostly from an
outflow which is not highly relativistic, but this does not in itself
rule out a highly relativistic jet being present in all objects,
perhaps forming a spine within a much slower outer sheath, as is also
required in the Blazars (Chiaberge et al. 2000;
Ghisellini, Tavecchio \& Chiaberge 2005; Chiaberge et al. 2005).
However, a BL Lac type spine emerging from every
AGN with $L/L_{Edd}<0.01$ is inconsistent with the statistics of the
GeV detections of AGN. The spine is highly beamed, so is aligned in
only 1 out of $\Gamma^2=225$ AGN, and the spine jet should scale with
accretion power (Heinz \& Sunyaev 2003). Nonetheless, the resulting
aligned jet is bright enough and the numbers of low luminosity AGN are
large enough that this over-predicts the observed Fermi All Sky Survey
numbers of BL Lacs by a factor of 1000 (Gardner \& Done 2014b).

Current consensus is that all jets are probably powered
by a combination of rotation and magnetic field, but there are two
main models for their formation, either using the rotational energy of
the black hole (Blandford \& Znajek 1977, hereafter BZ or spin powered
jet), or the accretion flow (Blandford \& Payne 1982, hereafter BP or
accretion powered jet). High black hole spin producing a spin powered
jet has long been identified as the possible trigger for launching and
powering the most highly relativistic jets (Blandford \& Znajek 1977;
Begelman, Blandford \& Rees 1984; Wilson \& Colbert
1995). This is supported by recent numerical simulations of the
accretion flow which show that truly relativistic outflows are produced
predominantly by spin power, while lower velocity outflows are driven
by a combination of BP and other mechanisms (see e.g. the review by
Yuan \& Narayan 2014). 

There are two types of jet in theory and numerical simulations, one
highly relativistic and one not, and two types of jet seen in
observations, one highly relativistic and one not. It is clearly
inherently a very attractive idea to associate the highly relativistic
jets with BZ and hence high spin black holes, and the rest to low spin
black holes producing only accretion powered outflows (e.g. Maraschi et
al. 2012). The only reason this very attractive idea was dropped was
the observational evidence for high black hole spin in radio quiet
objects (Ghisellini et al. 2004). However, the majority of high spin
measurements of radio-quiet objects are for complex NLS1 such as
1H0707 (see e.g. the compilation of Reynolds 2013). If these other
NLS1 are likewise super-Eddington, as claimed by Collin \& Kawaguchi
(2004), then these also have inner discs which are unlikely to be
flat, inconsistent with the lamppost/flat disc geometry used to derive
the high spin values. If these high spin values are not robust then
this removes the argument against the simplest models of jets, 
whereby
a low Lorentz factor accretion powered jet is present in all
sources, though its structure and power depends on the accretion flow
configuration, with a spin powered jet only in the highest spin
objects, again with its structure and power depending on the accretion flow
configuration. High spin can then be a necessary and sufficient condition
for a highly relativistic jet. 

If instead the `complex' NLS1 do indeed have high spin, then
the trigger for highly relativistic jets must instead be some
additional, external parameter such as accretion of magnetic flux from
the halo (Sikora \& Begelman 2013). Such models have no predictive
power, which limits the confidence with which the jet mode feedback
can be modelled in cosmological simulations of the growth of large
scale structure in the Universe (e.g. Bower et al. 2006).

\section{Conclusions}

Both 1H0707 and PG1244 have optical/UV spectra which clearly require
super-Eddington mass accretion rates through their outer disc.
However, these over-predict the observed bolometric luminosities so
there must also be advective and/or kinetic losses. The inner
accretion disc is then not at all likely to be flat, as assumed in the
lamppost models, with a smooth shape produced by pure advection losses
but with an additional clumpy turbulent wind produced in the numerical
simulations. Any such vertical structure on the disc means that
inclination angle is important, and can dramatically alter our view of
the central regions. We suggest that this is the origin of the
distinction between `simple' and `complex' NLS1, with the characteristic
deep X-ray intensity dips seen in the `complex' NLS1 being produced by
occultation of the central source by the upper edge of the vertical
structure.

We speculate that this may allow low spin models for the `complex' NLS1,
in which case high black hole spin may be a necessary and sufficient
condition required to produce highly relativistic jets.

\section*{Acknowledgements}
The authors thank the referee for their input which helped clarify the paper.
CD thanks Karen Leighly for the CTIO optical spectral data of 1H0707 shown
in Figure 1, and Xinlin Zhou for some earlier work on the X-ray spectra
of 1H0707. CD acknowledges STFC funding under grant ST/L00075X/1.
Funding for SDSS-III has been provided by the Alfred P. Sloan
Foundation, the Participating Institutions, the National Science
Foundation, and the U.S. Department of Energy Office of Science. The
SDSS-III web site is http://www.sdss3.org/.
This work is based on observations obtained with XMM-Newton, an ESA
science mission with instruments and contributions directly funded by
ESA Member States and the USA (NASA). All data used in this paper are
publicly available on the archives.


\label{lastpage}
\end{document}